\documentclass[showpacs,pra,aps,twocolumn]{revtex4}

\usepackage{bm}
\usepackage{amsmath}
\usepackage{amssymb}
\usepackage{latexsym}
\usepackage{amsfonts}
\usepackage{epsfig}
\usepackage{psfrag}
\usepackage{color}

\begin{document}

\title{Quantum dynamics and macroscopic quantum tunneling of two weakly coupled condensates}
\author{Ren\'e John Kerkdyk$^{1}$, and S. Sinha$^{2}$ }
\affiliation{$^{1}$Institut f\"ur Theoretische Physik, Georg-August-Universit\"at G\"ottingen, 37077 G\"ottingen, Germany\\
$^{2}$Indian Institute of Science Education and Research-Kolkata,Mohanpur, Nadia 741252, India}

\date{today}

\begin{abstract}
We study the quantum dynamics of a Bose Josephson junction(BJJ) made up of two coupled Bose-Einstein condensates. Apart from the usual ac Josephson oscillations, two different dynamical states of BJJ can be observed by tuning the inter-particle interaction strength, which are known as '$\pi$-oscillation' with relative phase $\pi$ between the condensates and 'macroscopic self-trapped' (MST) state with finite number imbalance. By choosing appropiate intial state we study above dynamical branches quantum mechanically and compare with classical dynamics. The stability region of the '$\pi$-oscillation' is separated from that of 'MST' state at a critical coupling strength. Also a significant change in the energy spectrum takes place above the critical coupling strength, and pairs of  (quasi)-degenerate excited states appear.
The original model of BJJ can be mapped on to a simple Hamiltonian describing quantum particle in an 'effective potential' with an effective Planck constant. Different dynamical states and degenerate excited states in the energy spectrum can be understood in this 'effective potential' approach.  
Also possible novel quantum phenomena like 'macroscopic quantum tunneling'(MQT) become evident from the simple picture of 'effective potential'. We study decay of metastable '$\pi$-oscillation' by MQT through potential barrier. The doubly degenerate excited states in the energy spectrum are associated with the classically degenerate MST states with equal and opposite number imbalance. We calculate the energy splitting between these quasi-degenerate excited states due to MQT of the condensate between classically degenerate MST states.

\end{abstract}


\maketitle

\section{Introduction}
The creation of Bose-Einstein condensate (BEC) of alkali atoms \cite{BEC,BEC1} has opened up various possibilities to study coherence properties of macroscopic 'matter wave'. One of such direction is to study the Josephson oscillation\cite{josephson} in two weakly coupled BECs. 
Similar to the superconducting Josephson junction, a Bose-Josephson junction (BJJ) can be created by two BECs in a double well potential, which are weakly coupled by the overlap of their wave functions\cite{stringari,javanainen}.  The condensates in each well are described by a 'matter wave' with a well defined phase. After they are weakly connected, the relative phase changes and a particle current is produced across the barrier. The nonlinear dynamics of ac and dc Josephson effects of BJJ have already been studied experimentally by loading condensates in a double well trap\cite{albiez,levy}. Also the matter wave interferometry have been studied in a double well geometry on an atom chip\cite{schumm}.
%The ac Josephson effect can be observed in coupled BECs either by tilting the trap slightly or by creating an initial number imbalance between the condensates.
The number difference between two Josephson coupled condensates and their relative phase are two canonically conjugate collective variables which show coherent oscillations when an initial number imbalance is created between the condensates. The non-linear dynamics of the relative phase and number imbalance between the coupled condensates have already been studied theoretically by several authors within the classical field approximation\cite{shenoy1,liang,sakellari,shenoy2}. 
%Like the superconductors, the dissipative effects in the BJJ are also present due to the presence of thermal atoms. 
Effect of dissipation in BJJ has been studied theoretically\cite{sols,shenoy3} and also been observed experimentally in the decay of Josephson oscillation\cite{levy}. Quantum fluctuations become
important for small number of particles in the condensate and for large interparticle interaction. Full quantum dynamics of BJJ with finite number of particle shows various interesting non-linear effects like 'revival' and 'beating' for different values of coupling constant\cite{walls,foerster}.

One of the main advantages of BJJ is the possibility to tune the interparticle interaction strength which enables to study various types of non-linear Josephson oscillations  by controlling the interaction strength. 
Apart from the usual ac Josephson oscillation two other types of dynamical branches can exist and their stability region is separated at a critical coupling strength. Above the critical coupling strength the 'macroscopic self-trapping'(MST) can occur dynamically with non vanishing time average of the population imbalance between the condensates\cite{shenoy1}. Below the critical coupling strength another dynamical state known as '$\pi$ -oscillation' can exist in which the relative phase between two condensates is $\pi$ \cite{shenoy1,shenoy2}. 
%Above a critical interaction strength a new branch of dynamical oscillation known as 'macroscopic self-trapped'(MST) state has been studied for which the time averaged value of number imbalance between the condensates is nonzero\cite{shenoy1,levy}.  
 
%In this paper, within single mode approximation the BJJ is described by two site Bose Hubbard model to study the quantum dynamics. By mapping the original Hamiltonian to a spin model we recover the 'pendulum model'\cite{shenoy1} describing the classical dynamics of the relative phase and number imbalance. 
We also notice a remarkable change in the energy spectrum associated with the change in dynamical branches, and doubly degenerate excited states appear above the critical coupling. The system of many-bosons in BJJ can be mapped on to a quantum mechanical problem describing a particle in an effective potential. 
The main focus of this work is to analyze different types of dynamical states and change in the energy spectrum in terms of this 'effective potential'.
We also address novel macroscopic quantum effects related to 'macroscopic quantum tunneling'(MQT) in this system which is evident from the method of 'effective potential'. The decay rate of
the metastable state corresponding to '$\pi$ -oscillation', and tunneling time between two MST states with equal and opposite number imbalance, are calculated analytically within semi-classical formalism of MQT. 
The MQT is always suppressed by the number of particles in the condensate, nevertheless this novel quantum effects can be observed in BJJ with small number of particles and close to the critical coupling strength.
%Also the doubly degenerate states in the energy spectrum can be understood from the effective potential approach. Apart from the Josephson oscillations, we consider the 'macroscopic quantum tunneling'(MQT) in this system. Below the critical coupling strength, the state of the condensate corresponding to the '$\pi$-oscillation' is metastable and it can decay by MQT through the potential barrier. 
%We calculate the decay rate semiclassically close to the critical coupling strength. Also above the critical coupling the appearance of doubly degenerate excited states can be understood in terms of a double well shaped effective potential whose minima corresponds to the fixed point of self-trapping with equal and opposite population imbalance. We consider the MQT of the system between these two MST states. The MQT is always suppressed by the number of particles in the condensate, nevertheless this novel quantum effects can be observed in BJJ with small number of particles and close to the critical coupling strength.

This paper is organized as follows. In section II, we model BJJ within single mode approximation by two site Bose-Hubbard model to study Josephson dynamics. By mapping the original Hamiltonian to a spin model we recover the 'pendulum model'\cite{shenoy1} describing the classical dynamics of the relative phase and number imbalance. Different dynamical branches and collective frequencies of small oscillation, obtained from the classical analysis are compared with the full quantum dynamics. The equivalence between the spin model of BJJ and the quantum mechanical problem describing a particle in an "effective potential" has been outlined in section III. Dynamical states below (and above) the critical coupling constant are analyzed from the effective potential. Next we outline possible novel collective quantum effects which emerge from the simple picture of 'effective potential'. In subsection A, we consider possible decay of meta-stable '$\pi$ -oscillation' due to MQT through the potential barrier. Subsection B is devoted to analyze the energy spectrum of the Hamiltonian and its connection to the self-trapping phenomenon. The appearance of doubly degenerate excited states in energy spectrum above the critical value of coupling can be understood in terms of a double well 'effective potential' whose minima corresponds to the dynamical fixed point of self-trapping with equal and opposite number imbalance. In subsection C we consider the MQT of the system between these two MST states and calculate the energy splitting between the quasi-degenerate excited states. Finally we summarize in section IV.

\section{The model and Josephson dynamics}   
Bose Josephson junction of two weakly coupled BECs can be well described by a two-site Bose-Hubbard model(BHM),
\begin{equation}
H = -\tilde{J}(a_{1}^{\dagger}a_{2} + a_{2}^{\dagger}a_{1}) + \frac{\tilde{U}}{2}\left[
\hat{n}_{1}(\hat{n}_{1} -1 ) + \hat{n}_{2}(\hat{n}_{2} -1)\right]
\label{jj_ham}
\end{equation}
where $a^{\dagger}_{i}(a_{i})$ is creation(annihilation) operator of bosons at site index $i$ ($i=1,2$ for two sites), $\hat{n}_{i} = a_{i}^{\dagger}a_{i}$ is number operator, $J$ is the tunneling matrix 
element, and $U$ is the on-site interaction strength. The tunneling matrix $J$ and the on-site interaction strength $U$ are the parameters of the effective model, which can be calculated from the exact shape of the trapping potential and macroscopic wave function of the condensates\cite{liang}.
For a fixed value of total number of particles $N$ in the condensates, we can transform the above Hamiltonian in Eq.(\ref{jj_ham}) to a spin model,
\begin{equation}
H = -JS_{x} + \frac{U}{2}S_{z}^{2},
\label{spin_ham}
\end{equation}
where, $J = 2\tilde{J}$, $U/2 = \tilde{U}$, $S_{x} = \frac{1}{2}(a^{\dagger}_{1}a_{2} + a^{\dagger}_{2}a_{1})$, $S_{y} = \frac{1}{2\imath}(a^{\dagger}_{1}a_{2} - a^{\dagger}_{2}a_{1})$, and $S_{z} = \frac{1}{2}(a^{\dagger}_{1}a_{1} - a^{\dagger}_{2}a_{2})$ are the Schwinger boson representation of a large spin of magnitude $S= N/2$.
In the effective spin Hamiltonian we neglect the constant term depending on the total number of particles $N$.

For a large spin of magnitude $S$, we can represent it classically by a spin vector
$\vec{S} = S(\sin\theta \cos\phi, \sin\theta \sin\phi, \cos\theta)$, where two angles $\theta$ and $\phi$ describes the orientation of the spin in the cartesian coordinate system.
Physically, the variables $\cos\theta$ and $\phi$ describe the fractional number imbalance and relative phase between two condensates in the original BJJ model. In terms of these variables the Lagrangian of the classical spin is given by,
\begin{equation}
\it{L} = S(\cos\theta)\dot{\phi} - H(\theta,\phi)
\label{spin_lag}
\end{equation}
where $\dot{\phi}$ is the time derivative of $\phi$, and $H(\theta,\phi) = -JS\sin\theta\cos\phi + \frac{US^2}{2} \cos^{2}\theta$ is the classical energy corresponding to the Hamiltonian in Eq.(\ref{jj_ham}). Here the variable $\cos\theta$ is canonically conjugate to the variable 
$\phi$ and plays the role of momentum. To use dimensionless variable, we scale time by $1/J$, energy by $J$ and introduce a dimensionless coupling constant 
$\alpha = \frac{US}{J}$, in order 
to make the total energy an extensive thermodynamical quantity.

From the Lagrangian given in Eq.(\ref{spin_lag}), we obtain the classical equations of motion of a large spin,
\begin{eqnarray}
& & \dot{\theta} = \sin \phi, 
\label{class_eqn1}\\
& & \sin \theta \dot{\phi} = \cos \theta \cos \phi + \alpha \cos \theta \sin \theta. 
\label{class_eqn2}
\end{eqnarray}
The linear stability analysis of the above dynamical equations reveals stable dynamics around three fixed points, given below:
\begin{itemize}
\item First fixed point is at $\theta = \pi/2$ and $\phi = 0$. It describes the ground state of the total system, which is the symmetric combination of the wave functions of two equally populated condensates in two well. The frequency $\omega$ of small amplitude oscillations around this fixed point is,
\begin{equation}
\omega/J = \sqrt{1 + \alpha}
\end{equation}
which is the plasmon frequency of ac Josephson effect. 
\item
A '$\pi$-{\it phase oscillation}' is described by the second fixed point at $\theta = \pi/2$, and $\phi = \pi$ \cite{shenoy1}. This fixed point represents the wave function of first excited state which is the anti-symmetric combination of wave functions of two condensates with equal number of atoms and relative phase $\pi$. The frequency of oscillation around this fixed point is given by $ \omega/J = \sqrt{1 - \alpha}$. This 
'$\pi$-oscillation' becomes dynamically unstable when $\alpha > 1$.
\item
A third fixed point at $\sin\theta = \frac{1}{\alpha}$, and $\phi = \pi$ represents a 'macroscopic self-trapped(MST)' state with non-zero population imbalance between two condensates\cite{shenoy1}. Classically this self-trapping phenomenon can be understood from a Ginzburg-Landau(GL) type potential. A second order differential equation for population imbalance $q = \cos\theta$ can be obtained from the Eq.(\ref{class_eqn1})and Eq.(\ref{class_eqn2}),
\begin{eqnarray}
& & \ddot{q} = -\frac{\partial V_{GL}(q)}{\partial q}\\
& & V_{GL}(q) = \frac{1}{2}(1 - E\alpha/S)q^2 + \frac{\alpha^{2}}{8}q^4
\label{eqn_gl}
\end{eqnarray}
where $\ddot{q}$ denotes second derivative of $q$ with respect to time, the conserved classical energy is given by $E/S = -\sqrt{1 - q^2}\cos\phi + \frac{\alpha}{2}q^{2}$. Above equation for number imbalance is exactly same as the classical equation of motion of a particle at $q$ in GL potential $V_{GL}(q)$\cite{polkovnikov}. This GL potential changes its shape when $E/S > \frac{1}{\alpha}$, and two minima appear for non vanishing population imbalance. For MST, the classical particle has total energy less than the barrier height in order to be localized at one of the minima of the GL potential. Two minima of the GL potential at population imbalance $ q = \pm \sqrt{1 - \frac{1}{\alpha^{2}}}$ correspond to the fixed points of MST states with energy $E/S = \frac{1}{2}(\alpha + 1/\alpha)$. The frequency of harmonic oscillation around the potential minimum is given by $\omega/J = \sqrt{\alpha^{2} - 1}$. For $\alpha > 1$, the GL potential can have two minima around which dynamically stable MST state can exist.
\end{itemize}

\begin{figure}
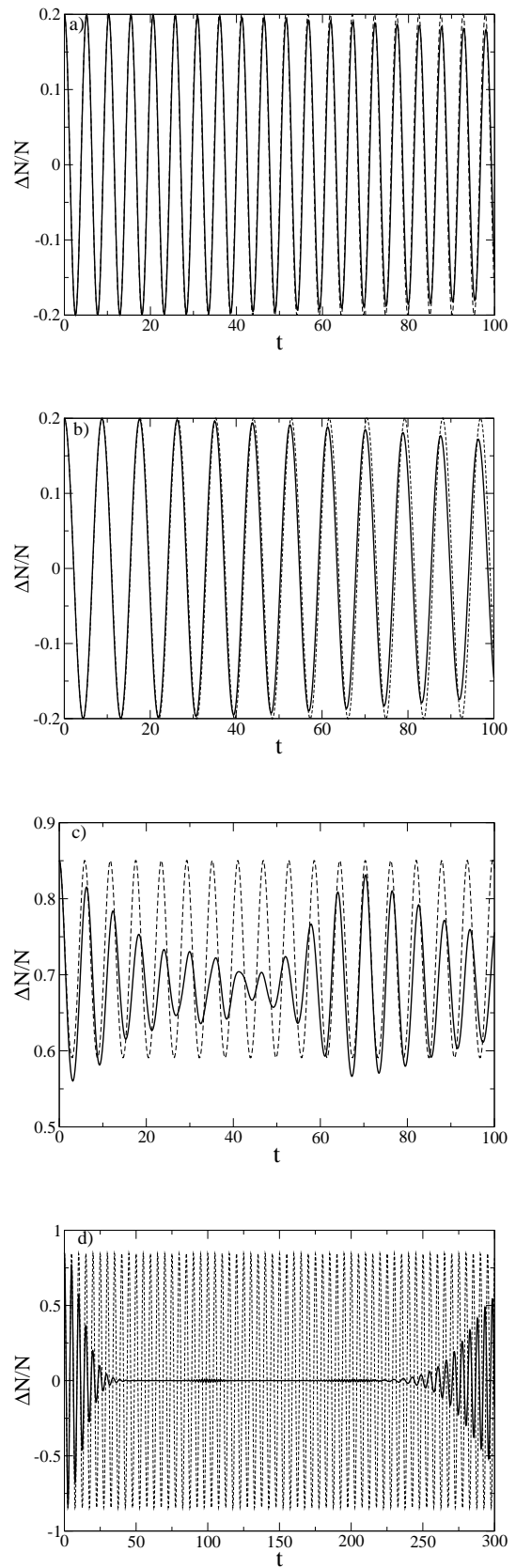

%\mbox[width=7.5cm]{
\begin{center}
\includegraphics[width=7cm]{fig1a.eps}
\end{center}
\vspace{.12cm}
\begin{center}
\includegraphics[width=7cm]{fig1b.eps}
\end{center}
\vspace{.12cm}
\begin{center}
\includegraphics[width=7cm]{fig1c.eps}
\end{center}
\vspace{.12cm}
\begin{center}
\includegraphics[width=7cm]{fig1d.eps}
\end{center}
%}
\caption{Oscillation of fractional population imbalance $\Delta N/N$ with time t (in units of $1/J$) for different values of coupling constant $\alpha$ and initial conditions. Results obtained from quantum dynamics(solid lines) are compared with classical dynamics (dotted line) for same initial conditions. a) Ac Josephson oscillation for $N=100$, $\alpha = 0.5$, and with classical initial condition $\cos \theta =0.2$, $\phi = 0$.
b)$\pi$-Oscillation for $N=100$, $\alpha = 0.5$, $\cos \theta = 0.2$, and $\phi =\pi$. c) Self trapping for $N=100$, $\alpha = 1.5$, $\cos \theta = 0.85$, and $\phi = \pi$. d) Damping and revival phenomena in quantum dynamics for $N=50$, $\alpha = 1.5$, $\cos \theta = 0.85$ and $\phi = \pi$.}
\label{fig1}
\end{figure}

From the above classical analysis we notice that $\alpha_c = 1$ is a critical coupling strength which separates two types of dynamical states. Next we
analyze the dynamical and spectral properties of the effective spin Hamiltonian (Eq.(\ref{spin_ham})) quantum mechanically.
In order to analyze various dynamical branches discussed above, we study numerically full quantum dynamics of BJJ with finite (but large) number of atoms and compare it with classical dynamics. We use the effective spin Hamiltonian (in Eq.(\ref{spin_ham})) to describe the quantum dynamics of BJJ.
Any quantum mechanical state of the system can be written as a linear combination of $2S + 1$ basis states,
\begin{equation}
|\Psi\rangle = \sum_{\sigma = -S}^{S} a_{\sigma} |\sigma\rangle,
\label{qstate}
\end{equation}
where $|\sigma\rangle$ is the eigensate of the operator $S_{z}$ with eigenvalue 
$\sigma$, and $a_{\sigma}$s are complex (in general time dependent) amplitude. To study the quantum dynamics, we time evolve any initial state by the time-evolution operator $e^{-\imath t H }$, and then evaluate the expectation values of the observables. To compare the quantum dynamics with the classical dynamics of the spin, we take the initial state of the spin as a coherent state corresponding to its classical orientation . Classical nature of a large spin is well described by a spin coherent state\cite{radcliff},
\begin{equation}
|\xi\rangle = \frac{1}{(1 + |\xi|^{2})^{S}} \sum_{\sigma=-S}^{S} \frac{\sqrt{(2S)!} \xi^{S-\sigma}}{\sqrt{(S + \sigma)!(S-\sigma)!}} |\sigma\rangle,
\label{coherent_st}
\end{equation}
where, the complex number $\xi = \tan\frac{\theta}{2}e^{\imath \phi}$ represents the orientation of the corresponding classical spin vector. 
We expand the initial state in terms of the eigen-vectors of the Hamiltonian and obtain the final state at time $t$, multiplying each component by a factor $e^{\imath E_{n}t}$ where $E_{n}$ is corresponding eigenvalue.
For $N=100$ ($S=N/2$), we calculate the time evolution of fractional population imbalance $(N1-N2)/N$ between two condensates (with average number of particles $N1$ and $N2$) in BJJ, from the expectation value of the operator $\hat{S}_{z}/S$, and compare it with the classical result obtained from time integration of Eqn.(\ref{class_eqn1}) and
Eqn.(\ref{class_eqn2}). To study different types of dynamics, we choose the initial values of the classical variables close to the fixed points and the initial coherent state can be obtained from the corresponding value of $\xi$. Three different types of Josephson dynamics (as discussed above) are shown in Fig.2, for different values of the coupling constant $\alpha$ and are compared with the classical dynamics. The ac Josephson dynamics and 
'$\pi$-oscillation' for $N=100$, show very good agreement with the classical dynamics. However the quantum dynamics around the MST state shows some deviation from the classical dynamics, also it shows some beating effect (see fig.1c). For the same initial number-imbalance but with phase $\phi=0$,quantum dynamics shows a damping and long time revival phenomena as shown 
in Fig1d.

\section{Effective potential method and macroscopic quantum tunneling}
In this section we consider interesting aspects related to the collective quantum effects in a BJJ in terms of the 'effective potential method' of a large spin\cite{zaslavskii}. The 'effective potential' of the quantum spin Hamiltonian gives a clear picture about various (meta)stable dynamical branches and also elucidate possible collective quantum tunneling in a BJJ.

First we outline the method of effective potential for the spin Hamiltonian in Eq.(\ref{spin_ham}) describing the BJJ.  
The eigenvalue equation of the Hamiltonian in Eq.(\ref{spin_ham}) can be written as,
\begin{eqnarray}
E a_{\sigma} &=& \frac{U}{2}\sigma^{2}a_{\sigma}  -\frac{J}{2}\big[\sqrt{(S + \sigma)(S - \sigma + 1)} a_{\sigma -1}\nonumber\\
& + &\sqrt{(S - \sigma)(S + \sigma + 1)} a_{\sigma + 1}\big],
\label{eigen_eqn}
\end{eqnarray}
where the eigenstates are expanded in terms of the basis states of $\hat{S}_{z}$, as described in Eq.(\ref{qstate}).
By introducing a generating function $\Phi(x) = \sum_{\sigma =-S}^{S} \frac{a_{\sigma} e^{\imath\sigma x}}{\sqrt{(S+\sigma)!(S-\sigma)!}}$ of an auxiliary variable $x$, above eigenvalue equation Eq.(\ref{eigen_eqn}) can be written as a differential equation \cite{zaslavskii},
\begin{equation}
E \Phi(x) = - \frac{U}{2}\frac{\partial^{2}\Phi}{\partial x^{2}} - \frac{J}{2}\left[2S\cos x \Phi(x) - 2 \sin x \frac{\partial \Phi}{\partial x}\right]
\end{equation}
It is interesting to note that $\Phi(x) = \langle\xi|\psi\rangle e^{\imath S \phi}$, with $\xi = e^{\imath x}$. So the variable $x$ can be interpreted as the phase $\phi$ between two condensates with equal number of particles in each trap i.e $|\xi| = 1$ (or $\cos\theta = 0$).
The transformation $\Phi(x) = e^{-J\cos x/U} \chi(x)$, reduces the above eigenvalue equation to an effective Schr\"odinger's equation,
\begin{equation}
{\cal{E}} \chi = - \frac{\alpha'}{2(s + 1/2)^{2}}\partial_{x}^{2}\chi + V(x)\chi
\label{schrodinger_eqn}
\end{equation}
where, ${\cal{E}} = E/(J(S + 1/2))$, $\alpha' = U(S + 1/2)/J \approx \alpha$, and the effective potential is given by,
\begin{equation}
V(x) = \frac{1}{2\alpha'} \sin^{2}x - \cos x.
\label{eff_pot1}
\end{equation}
First $2S + 1$ eigenvalues of the above Schr\"odinger's equation are the eigenvalues of the original spin Hamiltonian Eq.(\ref{spin_ham}). Above effective potential $V(x)$ for the relative phase can also be derived from the Eq.(\ref{class_eqn1})and Eq.(\ref{class_eqn2}) describing the classical dynamics. The relative phase $\phi$ between the condensates obeys the Newton's equation of motion in the effective potential $V(\phi)$ \cite{shenoy2}.
Above Hamiltonian in Eq.(\ref{schrodinger_eqn}) with the effective 
potential Eq.(\ref{eff_pot1}) is an exact quantum phase description (or 'rotor model') of Josephson junctions\cite{anglin}.

\begin{figure}
\centering
%\rotatebox{0}
\includegraphics[width=7cm]{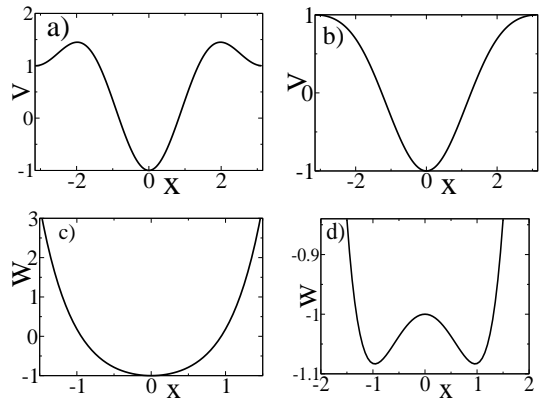}
\caption{Effective potential: $V(x)$ in the range $-\pi <x <\pi$, for a)$\alpha = 0.4$, and b) $\alpha = 1.4$. $W(x)$ for c)$\alpha =0.4$ and d) $\alpha = 1.4$.}
\label{fig2}
\end{figure}

It is interesting to note following significant points of the above equation:\\
\begin{itemize}
\item
The effective Planck constant in Eq.(\ref{schrodinger_eqn}) is $1/S$. Which implies that quantum fluctuation is suppressed for large spin systems (or large number of particles in the trap). 
\item
The role of effective mass is played by the inverse of the coupling constant $1/\alpha$. With increasing coupling constant $\alpha$ the fluctuation in number imbalance (or $S_{z}$) is reduced and the quantum fluctuation in phase $\phi$ is enhanced.
\item
As depicted in Fig.2a and Fig.2b, the shape of the effective potential $V(x)$ changes at a critical coupling $\alpha = 1$. For $\alpha < 1$, the potential has two minima at $x=2n\pi$ and $x = (2n +1)\pi$. These two minima correspond to the stable phase oscillations between the condensates discussed in previous section.
The $\pi$-oscillation corresponds to the local minima of the effective potential at $x = (2n+1)\pi$. The phase fluctuations around the classical steady state solutions can be obtained from harmonic approximation of the potential around the local minima. The potential at local minima $x_{\pm}$ can be approximated as, 
\begin{equation}
V(x) = {\mp} 1 + \frac{1}{2\alpha}(1 \pm \alpha)(x - x_{\pm})^{2}
\end{equation}
where, $x_{\pm}$ corresponds to the points $0$ and $\pm\pi$ respectively.
Note that the frequency of the harmonic potential exactly matches with the frequency of small amplitude oscillations. The phase fluctuations is given by,
\begin{equation}
\langle (x - x_{\pm})^{2}\rangle \approx \frac{\alpha}{2S \sqrt{1 \pm \alpha}}.
\label{phase_fluc}
\end{equation}  
Near the critical point the phase-fluctuation of $\pi$-oscillation becomes very large. 
\end{itemize}

\subsection{Decay of the '$\pi$-oscillation'}
As it is evident from the effective potential $V(x)$, the state with a relative phase difference $\pi$ between the condensates is a 
metastable state which is separated from the ground state by a potential barrier (see Fig.2a). A metastable state of many particles can decay by means of collective quantum tunneling through the barrier, known as 'macroscopic quantum tunneling'(MQT)\cite{coleman1}.
Since this collective state of many bosons is represented by a point particle in a meta-stable local minima of the effective potential $V(x)$, we can use simple quantum mechanical tunneling formula to calculate the collective decay rate.
The tunneling probability of a single quantum particle through a potential barrier scales as $\sim e^{-\sqrt{\frac{V_{max}}{m}}L/\hbar}$, where $V_{max}$ is the height of the potential barrier, L is the length of the barrier and $m$ is the mass of the particle.
In case of quantum phase model described by Eq.(\ref{schrodinger_eqn}), the effective Planck constant $\hbar$ scales by $1/S$ and hence the collective tunneling rate is suppressed by a factor of $S$ (or by the total number of particles $N$)in the exponential. As seen from the simple scaling analysis, the decay of the metastable '$\pi$-oscillation' state is suppressed exponentially by the number of particles $N$, and is negligible except in the region close to the critical coupling strength $\alpha \sim 1$ where the barrier height becomes vanishingly small. The potential near the meta stable state of $\pm \pi$-phase can be approximated as,
\begin{equation}
V(x) = 1 + \frac{\epsilon}{2}\tilde{x}^{2} -\frac{1}{8}\tilde{x}^{4} + ...
\label{approx_pot}
\end{equation}
where $\tilde{x} = x - x_{-}$, and $\epsilon = 1 -\alpha$ can be considered as a small parameter close to the critical point. To calculate the imaginary part of the ground state energy of the metastable state, we use the imaginary time path integral method. The imaginary time classical path in the quartic potential(in Eq.(\ref{approx_pot})) is known as instanton solution,
\begin{equation}
\tilde{x}(\tau) = 2 \sqrt{\epsilon} ~\mathrm{sech}\left[\sqrt{\epsilon}(\tau -\tau_{0})\right]
\label{instanton_soln}
\end{equation}
where $\tau$ denotes imaginary time. Above equation describes the classical path of a particle in the inverted potential given in Eq.(\ref{approx_pot}), where the particle starts from $x = x_{-}$ at $\tau = -\infty$, reaches the turning point $\tilde{x} = 2 \sqrt{\epsilon}$ at $\tau = \tau_{0}$, and returns to the original point at $\tau = \infty$. The imaginary time 'action' $S_{c}$ corresponding to this classical path is given by,
\begin{eqnarray}
S_{c} & = & \int_{-\infty}^{\infty} d\tau \left[(\frac{dx(\tau)}{d\tau})^{2} + V(x) - V(x_{-})\right], \nonumber\\
& = & \frac{16}{3} \epsilon^{3/2}
\label{im_action}
\end{eqnarray}
By using instanton technique, the decay rate $\Gamma$ of a quantum particle in an unstable quartic potential ( given by Eq.(\ref{approx_pot})) can be calculated in terms of the imaginary action $S_{c}$ and the frequency of oscillation $\omega$ at the metastable local minimum, and the analytical expression is $\Gamma = \omega \hbar \sqrt{6 S_{c}}{\pi \hbar}e^{- S_{c}/\hbar}$\cite{kleinert}. In the 'effective' Schr\"odinger equation approach of the spin model as written in Eq.(\ref{schrodinger_eqn}), $\hbar$ is replaced by $1/S$, and the frequency of oscillation at metastable state $x = \pm \pi$ is $\omega = \sqrt{1 - \alpha} \approx \sqrt{\epsilon}$. Substituting these parameters in the above expression, we obtain the decay rate of the metastable $\pi$-oscillation close to the critical point (when $\epsilon \ll 1$),
\begin{equation}
\Gamma = \sqrt{\frac{96}{3 \pi S} \epsilon^{5/2}} e^{-16 S \epsilon^{3/2}/3}.
\label{decay_rate}
\end{equation}

\subsection{Energy spectrum}
\begin{figure}
\centering
%\rotatebox{0}
%\begin{center}
\includegraphics[width=8cm]{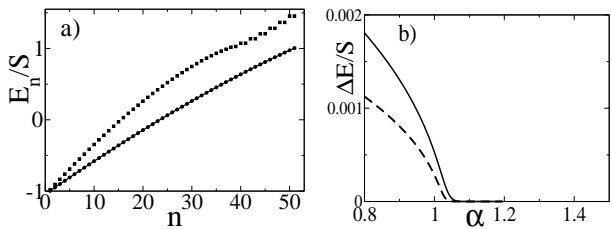}
%\includegraphics[width=4cm]{fig3b.eps}
%\end{center}
\caption{a)Energy eigenvalues of the Hamiltonian in Eq.(\ref{spin_ham}) for $N=50$, $\alpha = 0.4$(circles connected with line) and 
$\alpha = 2.5$(squares). b)Energy gap $\Delta E/S$ as a function of coupling strength $\alpha$ for $N=500$ (solid line) and $N=800$ (dotted line). }
\label{fig3}
\end{figure}

Similar to the change in shape of the potential $V(x)$, we also notice a distinct change in the energy spectrum of the spin Hamiltonian in Eq.(\ref{spin_ham}), as the coupling constant $\alpha$ is varied across the critical value $\alpha_c = 1$.  
We also analyze the energy spectrum of the Hamiltonian given in Eq.(\ref{spin_ham}) by numerical diagonalization, which is shown in the Fig.3 ( a and b) for $S=50$.
The interesting feature in the energy spectrum is appearance of doubly (quasi) degenerate excited states above the critical coupling strength $\alpha > 1$ (see Fig.3(b)).
First the degeneracy appears in the highest energy level, and then with the increasing coupling strength $\alpha$,
other excited states become pairwise degenerate. To quantify 
this transition in the energy spectrum, we calculate the energy gap $\Delta$ between the highest energy level $E_{2S + 1}$, and the next one $E_{2S}$ as a function of the coupling constant $\alpha$ (as depicted in Fig.2(c)). 
For finite S, the energy gap $\Delta$ vanishes slightly above the critical coupling $\alpha =1$. With the increasing values of S, the transition becomes sharper and the gap vanishes closer to the critical coupling.
This transition can be explained in terms of the 'effective potential' approach of the spin system and is directly linked to the 'self-trapping' phenomena. 

The energy spectrum of the original spin Hamiltonian in Eq.(\ref{spin_ham}) is exactly same as that of the Schr\"odinger's equation with an effective potential given in Eq.(\ref{schrodinger_eqn}), and Eq.(\ref{eff_pot1}). By performing two successive transformations, $x = x' + \pi$, and $x'= \imath y$, we obtain following eigenvalue equation from Eq.(\ref{schrodinger_eqn}),
\begin{equation}
\kappa \chi(y) = - \frac{\alpha'}{2(s + 1/2)^{2}}\partial_{y}^{2}\chi(y) + W(y)\chi(y),
\label{double_wpot}
\end{equation}
where, the effective potential $W(y) = \frac{1}{2\alpha} \sinh^{2} y - \cosh y$, and $\kappa = -{\cal{E}}$. Above transformation leads to the mapping of the highest eigenvalue $E_{2S+1}$ of the Schr$\ddot{o}$dinger equation given in Eq.(\ref{schrodinger_eqn}) to the ground state of the effective potential $W(y)$ in Eq.(\ref{double_wpot}). For $\alpha <1$, the effective potential has only one minimum and then it takes a shape of a double well for $\alpha >1$, as shown in Fig.2(c) and Fig.2(d).
Double well structure of the effective potential $W$ explains the appearance of doubly degenerate excited states of the original spin Hamiltonian for the coupling strength $\alpha >1$. When the kinetic energy term in the Hamiltonian (Eq.(\ref{double_wpot}) is neglected, the classical ground state energy
can be obtained from the potential $W$ at two degenerate minima. Classical ground state energy $\frac{1}{2}(\alpha + \frac{1}{\alpha})$ of the double well potential is exactly same as that of the fixed point corresponding to the MST state with population imbalance $\pm \sqrt{1 - 1/\alpha^{2}}$. As discussed earlier, the MST state can be described by the dynamics of a fictitious classical particle in the GL potential (in Eq.(\ref{eqn_gl})) with energy E less than the barrier height. The formation of the doubly degenerate excited states in the quantum mechanical spectrum can be understood from the periodic orbits of the classical motion near two minima of the GL potential. 

\subsection{Energy splitting and Macroscopic quantum tunneling between MST states }
For $S\rightarrow \infty$, the kinetic energy term in Eq.(\ref{double_wpot}) vanishes and two energy sates corresponding to the minima of the potential $W(y)$ are degenerate. But for finite values of $S$, the fictitious quantum particle described by Schr$\ddot{o}$dinger's equation in Eq.(\ref{double_wpot}), can tunnel between the minima of the potential and produce a small splitting between the highest degenerate eigenvalues discussed above.
%Next we consider the quantum effects due to the kinetic energy term in Eq.(\ref{double_wpot}) for finite values of spin S. A quantum particle in a double well potential can tunnel from one minima to other and produces a small splitting in degenerate energy states. 
The tunneling time T of the particle can be estimated from the relation $T\sim \hbar/\Delta E$, where $\Delta E$ is the energy splitting. This simple picture elucidates a novel phenomenon of MQT of the condensate in BJJ between the MST with equal classical energy. 

Next, we study the MQT of the condensate from one fixed point of MST with number imbalance $\sqrt{1 - \frac{1}{\alpha^{2}}}$ to the other fixed point $\sqrt{1 - 1/\alpha^{2}}$, which are represented by the minima of 'effective potential' $W$. 
The ground state energy splitting due to the quantum tunneling can be calculated by the instanton technique. The tunnel splitting of the ground state energy of the Schr$\ddot{o}$dinger's equation in Eq.(\ref{double_wpot}) has already been calculated by several authors in the context of uni-axial paramagnet \cite{enz,zaslavskii}. For $\alpha >1$, the effective potential develops two minima which satisfy the equation $\cosh(x_{min}) = \alpha$. The imaginary time classical path joining two minima of the potential can be calculated exactly
\cite{enz},
\begin{equation}
\mathrm{tanh} \frac{x}{2} = \sqrt{\frac{\alpha - 1}{\alpha + 1}} \mathrm{tanh}\frac{1}{2}\omega \tau,
\end{equation}
where $\omega = \sqrt{\alpha^{2} - 1}$ is the frequency of oscillation at the minima of the potential. Classical action corresponding to this instanton solution is given by,
\begin{equation}
S_c =  2 ln\left[\alpha + \sqrt{\alpha^{2} - 1}\right] - 2 \sqrt{\alpha^{2} - 1}/\alpha
\end{equation}
From the well known semiclassical formula for tunnel splitting\cite{coleman2,garg}, we obtain the energy difference between two MST states corresponding to the minima of the effective potential $W$,
\begin{equation}
\Delta E \approx 4 \sqrt{\frac{S}{\pi \alpha}} \frac{(\alpha^{2} - 1)^{2}}{\left[\alpha + \sqrt{\alpha^{2} - 1}\right]^{2S}}e^{2S\sqrt{\alpha^{2} - 1}/\alpha}.
\label{en_splitting}
\end{equation}
Apart from the non-linear oscillations around the self trapped state, the BJJ can show a very interesting quantum oscillation from one MST state to other MST state with equal and opposite number imbalance, and the estimated time period of such oscillation due to MQT is $\sim \frac{1}{S \Delta E}$. For large number of particle (or large S), the tunneling time will be very large and the system will
practically be trapped to one of the MST states depending on the initial condition.
But for finite number of particle in BJJ, and by tuning the coupling constant close to the critical point, such MQT can be observed.

\section{Summary}
To summarize, we consider two weakly coupled condensates described by an effective two-site BHM and studied quantum dynamics and collective quantum effects. 
We study quantum mechanically various dynamical branches of the Josephson oscillations and transition between them. Two-site BHM with N bosons can be mapped on to a simple model describing a quantum particle in an 'effective potential' where the Planck constant $\hbar$ is reduced by a factor of $N/2$. This 'effective potential' gives a clear picture of the '$\pi$-oscillation' and macroscopic self-trapping phenomena in BJJ. At the transition between these two dynamical branches, the 'effective potential' changes its shape.

Apart from the dynamics, the novel quantum effects like collective decay and MQT emerge from the simple picture of 'effective potential' method. The '$\pi$-oscillation' of the condensate  corresponds to a meta stable minimum of the potential $V(x)$, and can decay by means of MQT through the potential barrier. Semiclassically we calculate the decay rate $\Gamma$ for small potential barrier and obtain a power law behavior 
$\Gamma \sim (\alpha_c - \alpha)^{5/4}$ close to the critical coupling strength $\alpha_{c}$. 

We also notice a remarkable feature in the energy spectrum as the coupling constant $\alpha$ changes across the critical value. Quasi-degenerate pairs of excited states appear in the energy spectrum for coupling strength larger than the critical value $\alpha_{c}$.
The doubly degenerate highest energy states can be interpreted in terms of an effective double well potential $W(x)$ (shown in Fig.2(d)), and they correspond to the classically degenerate MST states with equal and opposite number imbalance. We study the MQT between these two MST states and estimate the tunneling time from the energy splitting. The MQT rate is exponentially suppressed  by number of particles N, nevertheless these novel macroscopic quantum phenomena can be observed in a BJJ with finite number of atoms and by tuning the coupling constant close to the critical value. 

%We study the dynamical branches of the Josephson oscillations, namely, the '$\pi$-oscillation' with a phase difference $\pi$ between the condensates and MST state with non vanishing number imbalance between the condensates. These two dynamical branches of Josephson dynamics are separated at a critical coupling strength where the spectral properties of the Hamiltonian also changes. We explain these meta stable states and 
%the change in energy spectrum by mapping BJJ to a quantum particle in an effective potential. Within this picture of effective potential we calculate the decay rate of meta stable $\pi {\it{oscillation}}$ state by using instanton calculus. Above the critical coupling strength, the appearance of doubly degenerate highest excited state can interpreted in terms of an effective double well potential and the degenerate states correspond to the MST with equal and opposite number imbalance. We also study the MQT of BJJ between these two MST states and estimate the tunneling time. For large number of particle, these quantum effects and MQT will be suppressed, nevertheless these novel macroscopic quantum phenomena can be observed for BJJ with finite number of atoms and by tuning the coupling constant close to the critical value. 

\section*{Acknowledgement}
We would like to thank P. A. Sreeram and A. M. Ghosh for helpful discussions.
RJK thanks the support of Georg-August-University G\"ottingen in the
framework of G-KOSS and IISER,Kolkata for local hospitality.

\end{document}